\documentstyle[12pt]{article}

\parskip=\medskipamount
\begin{document}
\baselineskip=20pt

\begin{center}
{\normalsize DISORIENTED CHIRAL CONDENSATE AND CHARGE-NEUTRAL PARTICLE
FLUCTUATIONS IN HEAVY ION COLLISIONS}\\
\vspace{1.5cm}
{ \normalsize  MLADEN MARTINIS, \footnote { e-mail address:
martinis@rudjer.irb.hr}
and VESNA MIKUTA-MARTINIS \footnote {e-mail address:
vmikuta@rudjer.irb.hr}
}\\
\vspace{0.5cm}
{\it Theoretical Physics Division \\
Rudjer Bo\v skovi\' c Institute, P.O.B. 1016, \\
41001 Zagreb,CROATIA \\}
\end{center}

{\Large \it Abstract}
\vspace{0.5cm}
\baselineskip=20pt

The posibility of large charge and isospin  fluctuations
in high-energy heavy ion-collisions is studied within the
framework of the nonlinear $\sigma $-model with quark degrees of
freedom. The multipion exchange potential between two quarks is
derived. It is shown that the soft chiral pion bremsstralung
leads to anomalously large fluctuations in the ratio of neutral to
charged pions, predicted by the formation of a disoriented chiral
condensate
(DCC). The factorization property of the scattering amplitude in the
impact
parameter space of the leading two-nucleon system
is used to study semiclassical production of pions in the central
region. We show that the DCC-type fluctuations are suppressed if a
large number of pions are produced in $\rho $-type clusters.
Our conclusion is supported through the calculation of two pion
correlation parameters
as a function of the $\rho $-to-$\pi $ ratio.

\vspace{1cm}

\newpage
\baselineskip=24pt
{\Large\it  1. Introduction}

 Central ultrarelativistic collisions at RHIC and LHC with more then 3000
 produced particles present remarkable opportunity to analyse
event-by-event fluctuations of
hadronic observables.  Such  single events analysis with large statistics
can
reveal new physical phenomena usually hidden  when averages over
a large statistical sample of events are made [1,2].
The numbers of particles produced in relativistic heavy-ion  collisions
can
differ dramatically from collision to collision due to the variation
of impact parameter (centrality dependence), energy deposition (leading
particle effect),
and other dynamical effects [2,3].
The fluctuations can also be influenced by novel phenomena such as
disoriented chiral condensate (DCC) [4,5].

 The best probes of such novel dynamics are fluctuations
of conserved quantities, because conservation
laws limit the degree to which final-state scattering can dissipate.
 Even globally conserved quantities such as energy,
net charge, isospin, baryon number and strangeness
can fluctuate when measured, e.g., in a limited phase space region.

Several methods have been proposed [3,7] to distinguish
between statistical and dynamic fluctuations . In high energy hadronic and
heavy-ion collisions the correlations of particle production are usually
analysed and described
in terms of  "short range correlations" (SRC) and "long range
correlations" (LRC).
The SRC  are mainly due to clustering of  outgoing particles. These
clusters may be hadronic
resonances which decay into a few particles or minijets [5,6,8].
The LRC  dominate in high multiplicity events and are mainly due to global
conservation laws.

The old puzzle in cosmic-ray observations is the existence of  few exotic
events characterized by an anomalously large
number of charged pions in comparison with the number of neutral pions,
the Centauros [9], indicating that there should exist a
strong negative long-range correlation between  two types of the pions.

Such long-range correlations are possible if pions are produced
semiclassicaly  and constrained by global conservation of isospin
[10--15].

Although the actual dynamical mechanism of the
production of a classical pion field in the course
of a high-energy collision is not known, there exist
numerous interesting recent theoretical attemps to explain Centauros
either as
different types of isospin fluctuations due to the  formation of a
disoriented chiral condensate (DCC) [16--20], or as multiparticle
Bose-Einstein
correlations (BEC) [21 ], or as the formation of a strange quark matter
(SQM) [22].
Among the most interesting speculations is the idea of DCC that localized
regions of misaligned chiral vacuum might occur during the
ultrahigh-energy hadronic and heavy-ion collisions
when the chiral symmetry is restored at high temperatures.
These regions, if produced, would behave as a pion laser, relaxing to the
ground state by coherent pion emission [ ]. It is  generally accepted that
the
fluctuation of the ratio of neutral to charged pions of the Centauro type
could be
a sign of the DCC formation provided that a single large domain is formed,
containing
a large number of low $p_{T}$ pions. Since the pions formed in the DCC are
essentially
classical they form a quantum superposition of coherent states with
different orientation in
isospin space. If all the pions in the domain are pointing in the same
isospin direction
and the condensate state is a pure isoscalar then the formation of DCC
leads to large
event-by-event fluctuations in the ratio  $f = n_{0}/n$ of the number
of $\pi _{0}$'s in the DCC divided by the total number of pions produced
in an event.
The probability distribution of $f$  inside  the DCC  domain is
[13,16,17,19]
\begin{equation}
P_{DCC}(f) = \frac{1}{2\sqrt{f}}
\end{equation}
 There are a variety of proposed mechanisms
other than DCC which also lead to the distribution (1) [23,24,25,26].
The distribution $ P_{DCC}(f)$ is different from the generic
binomial-distribution
expected in normal events which assumes equal probability for production
of
$\pi _{+}$, $\pi _{-}$ and  $\pi _{0}$ pions.
The emission of charged and neutral pions is then uncorrelated and  the
distribution
\begin{equation}
P_{B}(n_{0},n) = {n \choose n_{0}}\left(\frac{1}{3}\right)^{n_{0}}
\left(\frac{2}{3}\right)^{n -n_{0}}
\end{equation}
in the limit as $n\rightarrow \infty , n_{0}\rightarrow  \infty  $~
with  $ f$  fixed,  approaches a delta function at $f = 1/3$.

The possibility of  observing the DCC type fluctuations
critically depends on the size  and the energy content of the DCC domain.
If the domain is of the pion size,
the effect of DCC is too small to be observed experimentaly.
The early accelerator searches for Centauros and DCC  at CERN [27--30] and
at Fermilab [23,31] were thus unsuccesful.
 With the RHIC facility at BNL now, there is a possibility to consider
event-by-evevnt fluctuations of
 other  specific hadronic observables that might be more informative about
the  signal of DCC.
 There are various  factors that may affect the observation of the DCC
signal [32].
Experimental signals such as the isospin fluctuations,
the strong relative enhancement of the number of low $p_{T}$ pions,
and the suppresion of HBS correlations may provide a robust signal of DCC
[33].

The space-time scenario of the formation and decay of the DCC is usually
studied within
one of the simplified versions of the chiral effective Lagrangians, either
the linear or nonlinear
sigma model [20]. However, it should be emphasized that the use of $\sigma
$- models,
be they linear or nonlinear, is only a rough approximation to the true
dynamics, because the
couplings of pion and sigma  fields to the constituent quarks may
be large and their effect should not be ignored.

In this paper, we present  results of  an event-by-event  analysis of
charge-neutral
pion fluctuations as a function of the $\rho /\pi $ ratio in
pp-collisions.
Following the approach of our earlier papers [24,34],
 we consider in Section 2. the leading-particle effect as a  source
of a classical pion field in the impact parameter space which is related
to the semiclassical solutions of the
equation of motion of the  nonlinear $\sigma $- model coupled to quark
degrees of freedom.
In order to faciliate the analysis of charged-neutral pion correlations,
we also derive the corresponding pion-generating function.
The quantum equations  of the nonlinear $\sigma $-model coupled to quarks
are discussed
in Section 3., and shown to lead to a coherent state description of the
pions emerging from the DCC.
A multipion exchange potential between two  quarks is derived as well as
a corresponding correlator for  multipion production.
Results of our investigation are summarized in the Section  4. Our general
conclusion is
that within the nonlinear $\sigma $ model the large isospin
fluctuations depend strongly on the value of the $\rho /\pi $ ratio which
fluctuate from event to event.

{\Large \it  2. Pion production from a classical  source}

At high energies most of the pions are
produced in the nearly baryon-free central region.
The energy available for the hadron production is
\begin{equation}
E_{had} = \frac{1}{2} \sqrt{s} - E_{leading}
\end{equation}
which at fixed total c.m. energy $ \sqrt{s} $ varies from event to event.
The n-pion contribution to the $s$-channel unitarity can be written as an
integral over the relative impact parameter  $b$ of the two
incident leading particles:
\begin{equation}
F_{n}(s) = \frac{1}{4s} \int d^{2}b \prod_{i=1}^{n}dq_{i}
\mid T_{n}(s, \vec{b};1 \ldots n) \mid^{2},
\end{equation}
where $ dq = d^{2}q_{T}dy/2(2 \pi)^{3}.$ The normalization is
such that
\begin{eqnarray}
F_{n}(s) & = & s \sigma_{n}(s), \nonumber \\
\sigma_{inel}(s) & = & \sum_{n=1}^{ \infty} \sigma_{n}(s).
\end{eqnarray}
If the isospin  of the incoming
leading particles is $II_{3}$ , then the initial-state
vector of the pion field is $ \hat{S}(s, \vec{b}) \mid II_{3} \rangle,$
where $ \mid II_{3} \rangle$ denotes a vacuum state with no pions.
The $n$-pion production amplitude is
\begin{equation}
iT_{n}(s, \vec{b};q_{1} \ldots q_{n}) = 2s \langle I'I'_{3};q_{1} \ldots
q_{n} \mid \hat{S}(s, \vec{b}) \mid II_{3} \rangle,
\end{equation}
where $ I'I'_{3}$ denotes isostate of the outgoing leading particles.

The basic assumption of the independent emission of  pion-clusters,
in b-space is the factorization property of the scattering amplitude.
The factorization of $T_{n}$ follows if
the quantum field  of the pion-cluster satisfies the  equation of motion
of the form
\begin{equation}
( \Box + m_{c}^{2}) \vec{ \pi _{c}}( s,\vec{b}; x) = \vec{j_{c}}(s,
\vec{b};x),
\end{equation}
for an isovector cluster and similar one for an isoscalar cluster.
Here, $ \vec{j_{c}}$ is a classical source of the cluster which decay
into $c$ = 1,2,... pions outside the region of strong interactions (the
final-state
interaction between pions being neglected).
Clusters decaying into two or more pions simulate a short-range
correlation
between pions. They need not be well-defined resonances.

If the conservation of
isospin is a global property of the colliding system,
then $ \vec{j_{c}}(s, \vec{b};x)$ is of the form
\begin{equation}
\vec{j_{c}}(s, \vec{b};x) = j_{c}(s, \vec{b};x) \vec{n},
\end{equation}
where $ \vec{n}$ is a fixed unit vector in isospace independent of $x, s$
and $\vec{b}$ .
The global conservation of isospin thus introduces the long-range
correlation
between the emitted pions.

The $S$ matrix following from such a classical source
is still an operator in the space of pions.
Inclusion of isospin requires $ \hat{S}(s, \vec{b})$
to be also a matrix in the isospace of the leading particles.

The coherent production of pion-clusters
is described by the following $S$ matrix:
\begin{equation}
\hat{S}(s, \vec{b}) = \int d^{2} \vec{n} \mid \vec{n}
\rangle \hat{D}( s, \vec{b}) \langle \vec{n} \mid,
\end{equation}
where $ \mid \vec{n} \, \rangle $ represents the isospin-state
vector of the two-leading-particle system.
The quantity $\hat{D}( s, \vec{b})$ is the unitary coherent-state
displacement operator defined in our case as
\begin{equation}
D( s, \vec{b}) = exp [a^{\dagger}(s, \vec{b}) - a(s, \vec{b})]
\end{equation}
where
\begin{equation}
 a^{\dagger}(s, \vec{b}) =  \sum_{c }\int dq [
J_{c}(s, \vec{b};q) \vec{n} \vec{a_{c}}^{ \dagger}(q) + J_{c}'(s,
\vec{b};q) a_{c}^{\dagger}(q)],
\end{equation}
and $ \vec{a_{c}}^{ \dagger}(q)$ and $ a_{c}^{ \dagger}(q)$ are
the creation operators of a
 cluster of type c, respectively.

The isospin $(I', I_{3}')$ of the outgoing leading particle
system varies from event to event.It is  produced with the
 probability $\omega_{I', I_{3}'}$, and we can sum over all $(I'I'_{3})$
 to obtain the  probability distribution of producing $n_{+} \pi^{+}, \,
n_{ \_} \pi^{ \_},$ and $n_{0} \pi^{0}$ from a given isospin state:
\begin{eqnarray}
P_{I I_{3}}(n_{+}n_{ \_}n_{0}) N_{I I_{3}} & \! \! = \! \! &
\sum_{I'I'_{3}}\omega_{I', I_{3}'} \int d^{2}bdq_{1}dq_{2} \ldots
dq_{n} \mid \langle I'I'_{3}n_{+}n_{ \_}n_{0} \mid \hat{S}(s,
\vec{b}) \mid II_{3} \rangle \mid^{2}, \nonumber \\
 n & \! \! = \! \! & n_{+} + n_{ \_} + n_{0}
\end{eqnarray}
where $ N_{I I_{3}}$ is the corresponding normalization factor determined
by \\
$ \sum_{\{n\}} P_{I I_{3}}(n_{+}, n_{-}, n_{0}) = 1 $.

This is our basic relation for calculating various pion-multiplicity
distributions, pion multiplicities, and pion
correlations between definite charge combinations.
In general, the probability $P_{I I_{3}}(n_{+}n_{ \_}n_{0})$
would depend on $(I'I_{3}')$ dynamically. The final-leading-particles
 tend to favor the $(I',I_{3}') \approx (I,I_{3})$ case.
 However, if the leading particles are colliding nuclei, an almost equal
probability for various $(I',I_{3}')$ seems reasonable aproximation
owing to the large number of possible leading isobars in the final state.
Then we can sum over all $(I'I'_{3})$ using the group theory alone.

Recent studies of heavy-ion collisions at the partonic level [26]
argue that the central region is mainly dominated by gluon jets. The
valence quarks of the incoming particles which
escape from the interaction region form the outgoing leading particle
system. Since gluon's isospin is zero, it is very likely that total
isospin of the produced pions in the central region is also zero. This
picture is certainly true if the central region is free from valence
quarks, the situation expected to appear at the extremelly high collision
energies.

Let us now assume that pions are produced both singly and through
isovector
clusters of the $ \rho$ type [24]. In this case, the most appropriate tool
for
 studying various pion correlations is the generating function
$G_{II_{3}}(z,n_{ \_})$:
\begin{equation}
G_{II_{3}}(z,n_{ \_}) = \sum_{n_{0},n_{+}}P_{II_{3}}(n_{+},n_{ \_},
n_{0})z^{n_{0}},
\end{equation}
from which we can calculate, for example
\begin{eqnarray}
\langle n_{0} \rangle_{n_{ \_}} & = & \frac{d}{dz} \ln G_{II_{3}}
(1,n_{ \_}), \\
f_{2,n_{ \_}}^{0} & = & \frac{d^{2}}{dz^{2}} \ln G_{II_{3}}(1,n_{ \_}),
\end{eqnarray}
and
\begin{equation}
P_{II_{3}}(n_{0}) = \frac{1}{n_{0}!}\frac{ d^{n_{0}}}{dz^{n_{0}}}
\sum_{n_{ \_}}
G_{II_{3}}(0, n_{ \_}). \\
\end{equation}
The form of this generating function is
\begin{equation}
G_{II_{3}}(z,n_{ \_})  =  (I + \frac{1}{2})
 \frac{(I-I_{3})!}{(I+I_{3})!}
\int_{-1}^{1}dx \mid P_{I}^{I_{3}}(x) \mid^{2} \frac{ \textstyle
A(z,x)^{n_{0}}}{ \textstyle n_{0}!}e^{ \textstyle - B(z,x)},
\end{equation}
where
\begin{equation}
2A(z,x) = (1-x^{2}) \overline{n}_{ \pi}+z(1-x^{2}) \overline{n}_{ \rho}+
2x^{2} \overline{n}_{ \rho}
\end{equation}
and
\begin{equation}
2B(z,x) = \overline{n}_{ \pi}(1+x^{2} - 2zx^{2}) +
 \overline{n}_{ \rho} (2 - z(1-x^{2})).
\end{equation}
Here $ \overline{n}_{ \pi}$ denotes the average number of directly
produced pions, and $ \overline{n}_{ \rho}$ denotes the average number
of $ \rho$-type clusters which decay into two short-range correlated
pions. The function $P_{I}^{I_{3}}(x)$ denotes the associate
Legendre polinomial. Note that $A(1,x)=B(1,x)$.

The total number of emitted pions is
\begin{equation}
\overline n = \overline{n}_{ \pi} + 2 \overline{n}_{ \rho}.
\end{equation}

The behavior of $P (n_{0})\equiv P_{00}(n_{0})$ for
$ \overline n = 50$ and different combinations of $( \overline{n}_{ \pi},
\overline{n}_{ \rho}) $ is shown in Fig. 1.The behaviour of the
$n_{0}$-dispersion
\begin{equation}
D(n_{0})_{n_{ \_}}^{2} =  f_{2,n_{ \_}} - \langle n_{0} \rangle _{n_{ \_}}
\end{equation}
 for a given number of negative pions and  different pairs of
$( \overline{n}_{ \pi},\overline{n}_{ \rho})$ in the case of $I=I_{3}=1$
is shown in Fig. 2..
Note that $f_{2,n_{ \_}}$  is
a sensitive quatity of the pairing properties of the pions.

We see that DCC-type behavior is expected only for $ \overline{n}_{ \pi}
\neq 0$ and
$ \overline{n}_{ \rho}=0.$  that is in events without $\rho $-resonances.
Recent estimate of the ratio of
$ \rho$-- mesons to pions, at accelerator energies, is
$ \overline{n}_{ \rho} = 0.10  \overline{n}_{ \pi}$ [6].
\baselineskip=24pt

{\Large \it 3. DCC from a quantum nonlinear sigma model}

In this section we establish the relationship  between the quantum
nonlinear sigma model coupled to quarks and
our cohetrent state eikonal model description of the pions emerging from
the DCC.

As is well known, the Lagrangian for QCD with two light up and down quarks
has an approximate global
$SU(2)_{L}\times SU(2)_{R} $ symmetry, which at low temperatures, is
spontaneously broken
to $SU(2)_{V}$ by a nonzero value of the quark condensate $\langle
\bar{q}_{L}q_{R}\rangle $,
which is regarded as an order parameter of the system. This order
parameter
can be represented as a four-component vector $\phi \equiv ( \sigma ,
\vec{\pi })$ buildt from the quark densities.The
chiral symmetry then corresponds to $O($4$)$ rotations in internal space.

The true vacuum of the theory is defined as $ \langle \phi \rangle = (
\langle \sigma \rangle, \vec {0}) $,
with $ \langle \sigma \rangle \neq 0 $.
In QCD the spontaneous symmetry breakdown leads to nearly massless
Goldstone bosons (the pions) and gives the constituent-quark mass.
At low energies and  large distances ( momentum scale smaller than $ 1 GeV
$ ) the dynamics of QCD
is described by an effective Lagrangian containing the $\sigma , \vec{\pi}
$ fields and constituent quarks.

In the DCC dynamics we distinguish three stages: formation, evolution and
decay stage.
In the conventional approach [19] one starts with a chirally symmetric
phase at $ T T_{c} $ and
DCC formation happens as T, due to a rapid expansion or cooling, drops
below $ T_{c} $
spontaneously breaking the chiral symmetry.

The evolutionary stage of the DCC is usually described by the classical
chiral dynamics
based on the $ \sigma $-model, mostly the linear $ \sigma $- model.
We consider the nonlinear
$\sigma$-model coupled to quarks at zero temperature [35] which is
expected to describe the late stage of the DCC evolution.

The Lagrangian for the nonlinear $\sigma$-model coupled to quarks is
\begin{equation}
L = \frac{f_{\pi}^{2}}{4} Tr (\partial_{\mu}U^{\dagger} \partial^{\mu} U )
+ \overline{q} ( i \gamma \partial) q - g f_{\pi} \overline{q} U q \\
\end{equation}
where
\begin{equation}
U = exp ( i \gamma_{5} \frac{\vec{\pi} \cdot \vec{\tau}}{f_{\pi}}) \\
\end{equation}

We parametrize the pion field in the following form
\begin{equation}
\vec{\pi}(x) = f_{\pi} \vec{n}(x) \theta(x) \\
\end{equation}
where $ \vec{n}(x) $ is an unit vector which determines the isospin
orientation of the pion field, obeying $ {\vec{n}}^{2} = 1 $.

The Euler-Lagrange equations of motion for $ \theta $ and $ \vec{n} $ are
\begin{eqnarray}
\Box \theta - \sin{\theta}\cos{\theta} \partial_{\mu}\vec{n}\cdot
\partial^{\mu}\vec{n} & = & -i \frac{m_{Q}}{f_{\pi}^{2}} \vec{n} \cdot
(\bar{Q} \vec{\tau} \gamma_{5} Q ) \nonumber \\
\partial_{\mu}( {\sin}^{2}{\theta} \vec{n}\times \partial^{\mu} \vec{n} )
& = & -i \frac{m_{Q}}{f_{\pi}^{2}} \vec{n} \times (\bar{Q} \vec{\tau}
\gamma_{5} Q )\sin {\theta} \\
\end{eqnarray}
where Q denotes the constituent quark defined by
\begin{equation}
Q = e^{i\gamma_{5} \frac {\vec{\pi} \cdot \vec{\tau}}{2 f_{\pi}}} q \\
\end{equation}
and $ m_{Q} = g f_{\pi} $ is the constituent quark mass.

We treat
\begin{equation}
-i \frac{m_{Q}}{f_{\pi}} \bar{Q} \vec{\tau} \gamma_{5} Q = \vec{j}(x) \\
\end{equation}
as a given classical external source of pions and identify it with the
source function $\vec{j}_{1}(s,\vec{b};x)$
in our eikonal model.
For the class of solutions that can be rotated into a uniform one, $
\vec{n}(x) = \vec{n}$,
known as the Anselm-class of solutions [16], the solutions with constant $
\vec{n} $ can be realized if
the source points to a certain fixed direction $ \vec{n} $ in the
isospace:
\begin{equation}
\vec{j}(x) = j(x) \vec{n} \\
\end{equation}

Then the equation of motion for the pion field reduces to
\begin{equation}
\Box \theta (x) = j(x) \\
\end{equation}
with
\begin{equation}
\vec{\pi}(x) = f_{\pi}  \theta(x) \vec{n} \\
\end{equation}

This relationship offers us the possibility to study the importance of
various quark sources in the  DCC formation.

 In quantum chiral field theory  of the nonlinear   $\sigma $-model the
role
of a strong coupling of  $q\pi $-interaction has the quantity
$m_{Q}/f_{\pi }$.
 Since the Lagrangian of the nonlinear
  $\sigma $-model  is nonpolynomial, a suitable renormalization procedure
and
operator normal ordering should be formulated [29].
 Let
\begin{equation}
\Phi = :  e^{i\gamma _{5}\frac{\vec{\pi}\vec{\tau}}{f_{\pi}}} - 1 :
\end{equation}
denote the chiral super field of the pion. The Lagrangian describing the
interaction of
this super chiral field with quarks is now
\begin{equation}
L  =  \frac{f_{\pi}^{2}}{4}Tr(\partial _{\mu}\Phi \partial ^{\mu}\Phi ) +
\bar{q}( i \gamma \partial  -  m_{Q} )q  -
\frac{m_{Q}}{f_{\pi}}\bar{q}\Phi q .
\end{equation}
The chiral super propagator  of the field $\Phi $ is defined by
\begin{equation}
\bigtriangleup _{\Phi }(x) = \langle T(\Phi (x)\Phi (0))\rangle
\end{equation}
Its explicite form, after number of algebraic manipulations, is
\begin{equation}
\bigtriangleup_{\Phi }(x) =  \{ 1\otimes 1 + \frac{f_{\pi}^{2}}{3}
(\gamma _{5}\vec{\tau})\otimes (\gamma _{5}\vec{\tau}) \partial
_{\triangle}\}
\partial _{\triangle}[\triangle ch(\frac{\triangle }{f _{\pi}^{2}})] + 1
\end{equation}
where
\begin{equation}
\langle T(\pi _{i}(x)\pi _{j}(0))\rangle  =  \delta _{ij} \triangle (x).
\end{equation}
and \\
\[ \triangle (x) =  \frac{1}{4\pi ^{2}}\frac{1}{x^{2} - i\epsilon }\]. \\
The multipion exchange potential between two quarks is related to the
Fourier transform of $\triangle _{\Phi }(x)$ in the following way:
\begin{equation}
\bar{u}(p_{1}')\bar{u}(p_{2}') \tilde{\triangle }_{\Phi }(q)
u(p_{1})u(p_{2}) =
\omega '_{1}\omega '_{2} \tilde{V}_{\Phi }(q)\omega _{1}\omega _{2}
\end{equation}
where \\
\[ u(p) = \sqrt{\frac{\epsilon _{p} + m}{\epsilon _{p}}}
{\omega \choose \omega '}\] \\
and \\
\[ \omega ' = \frac{\vec{\sigma }\vec{p}}{\epsilon _{p} + m}\omega \] \\
with $\omega ^{*}\omega  = 1$.

For studying  the multipion production on quarks, it is necessary to find
the correlator
\begin{equation}
\bigtriangleup_{nm}(q_{1}, q_{2},...,q_{n};l_{1}, l_{2},...,l_{m}\mid x) =
\langle n\mid T(\Phi (x)\Phi (0))\mid m\rangle
\end{equation}
This is an involved task and has not been done yet for arbitrary n and m.

In the case of soft chiral pion bremsstralung [37-40],
in which every incoming and outgoing quark line is replaced by
\begin{equation}
q \longrightarrow  exp (i\gamma _{5}\frac{\vec{\pi }\vec{\tau }}{2f
_{\pi }}) q
\end{equation}
the distribution of neutral pions for large n is of the form
$1/\sqrt{n_{0}n}$ which
is typical for coherent pion production without invoking the notion of DCC
formation.

Therefore, we  conclude  that  $n P(n_{0}\mid n) \sim \sqrt{n/n_{0}}$
behaviour is not always a definite
signature of DCC formation.

{\Large \it  4.Conclusion}

The results of the present analysis have shown  the experimental
observation of DCC is strongly
affected by the $\rho$/$\pi$ ratio. In particular, we have found that:
\begin{itemize}
     \item Within the framework of an unitary eikonal model with
               factorization, energy conservation  and global conservation
of isospin the DCC-type fluctuation
           of the neutral pion fraction ($f$) could  be observed   if the
$\rho$/$\pi$ ratio  is small in an event;
     \item  The DCC effect also depends on isospin of the initial-leading-
                particle system;
      \item  The coherent production of  $\rho $-type clusters supresses
the
                 DCC-type behaviour;
       \item  The factorization property of
                  the scattering amplitude in the impact parameter space
may
                  be related to the isospin-uniform solutions of the
quantum nonlinear
                  $\sigma $-model coupled to quarks;
       \item   The multipion exchange potential between two quarks can be
found;
        \item   The soft chiral pion bremsstralung  also leads to
anomalously
                    large  ratio of neutral to charged pions.
\end{itemize}

\baselineskip=24pt
{\large \bf Acknowledgment }
This work was supported by the Ministry of Science of
Croatia under Contract No. 0098004.

\newpage

{\bf Figure captions :}

Fig.1. The curves represent $ P (n_{0}) $ for different combinations of
$( \overline{n}_{ \pi},  \overline{n}_{ \rho}),$
the average number of singly produced pions and the average
number of $ \rho$-type clusters, respectively.

Fig. 2. The dispersion  for two neutral pions as a function
of the number of negative pions  for $I = I_{3} =
1.$ The curves represent different combinations of
$( \overline{n}_{ \pi}, \overline{n}_{ \rho}).$

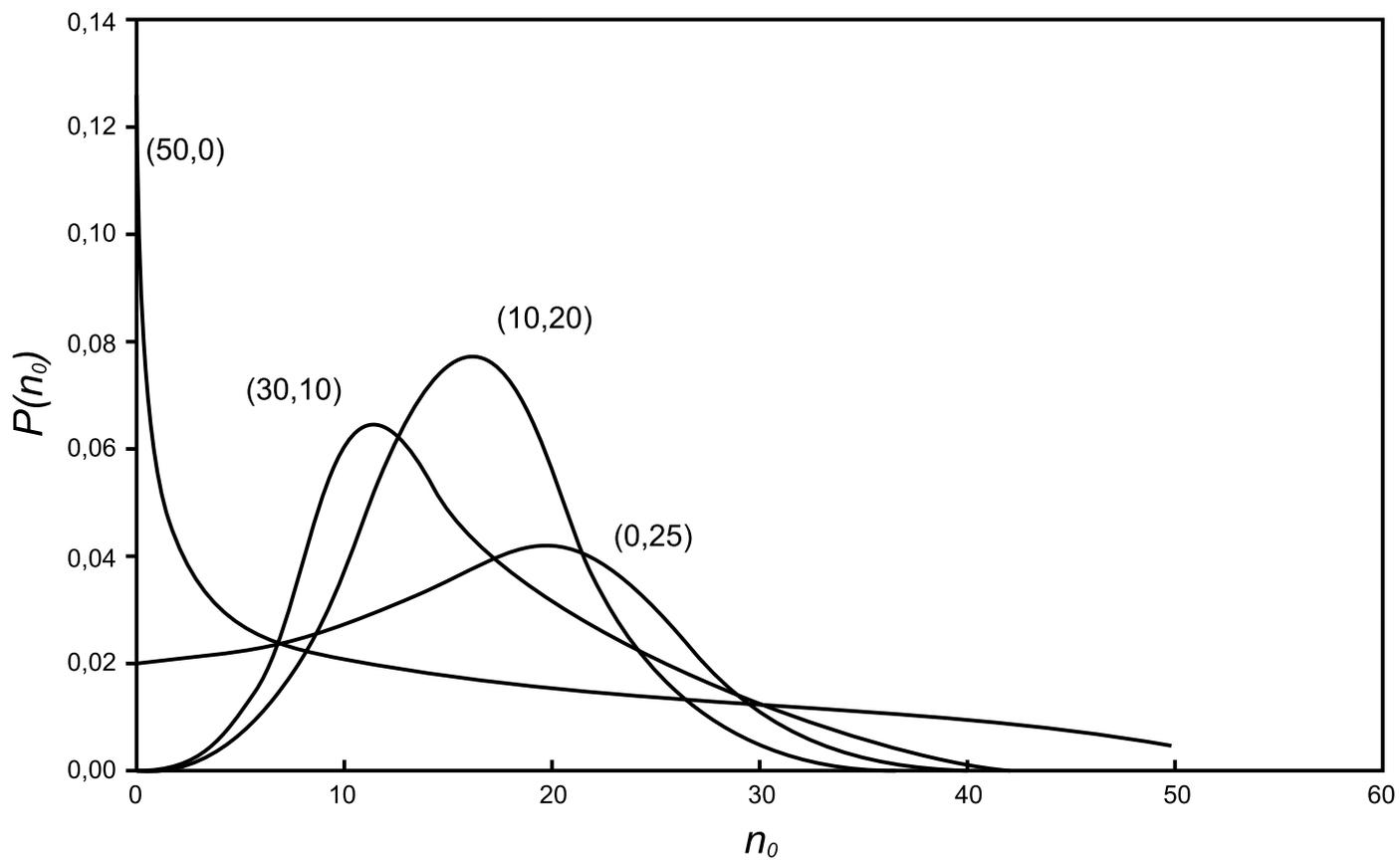
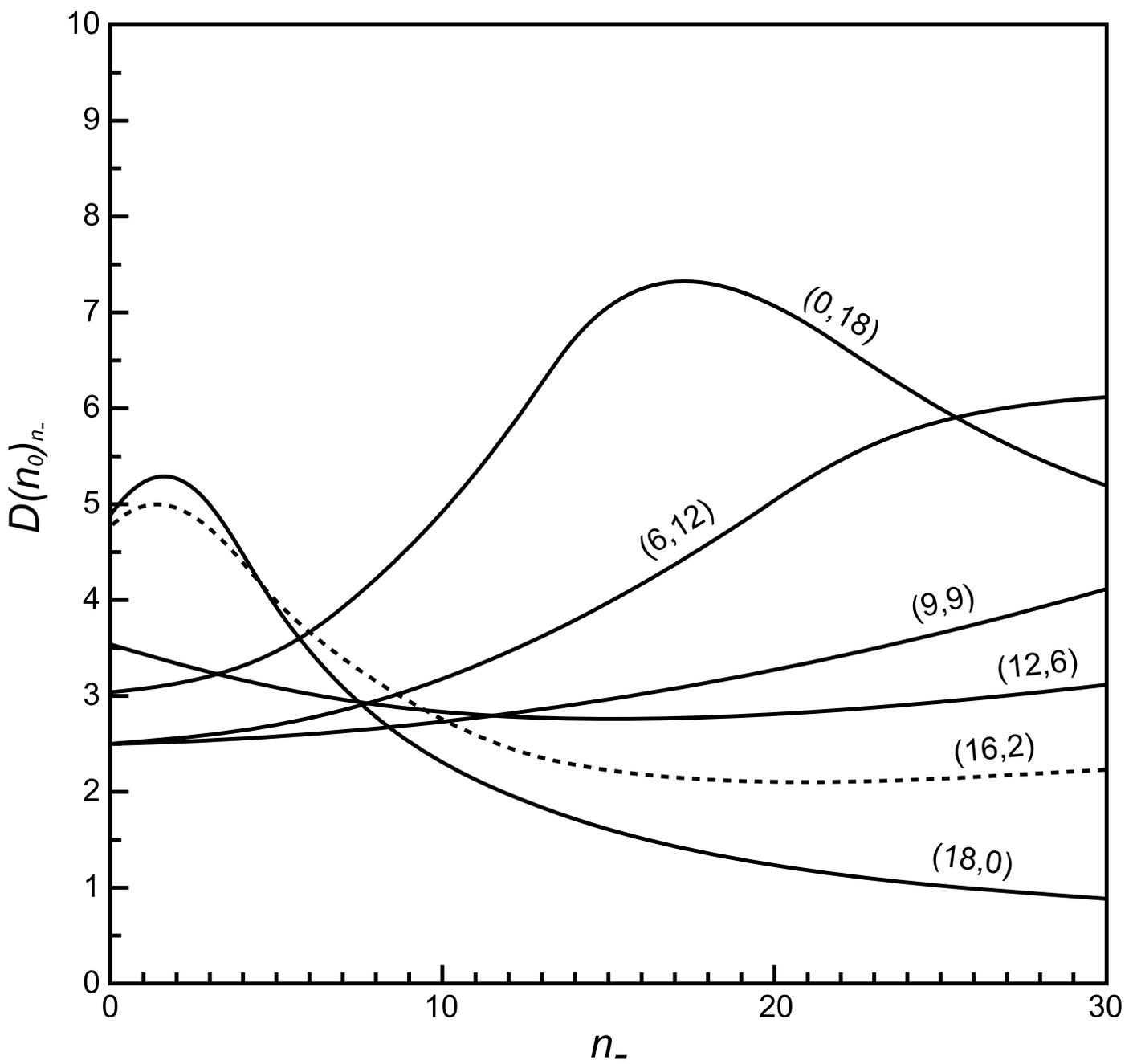
\end{document}